\documentclass{PoS}

\title{Radio source evolution and the interplay with the host galaxy}

\ShortTitle{Radio source evolution}

\author{\speaker{Manel Perucho}\thanks{This work has been supported by the Spanish Ministerio de Econom\'{\i}a y Competitividad (grants AYA2015-66899-C2-1-P and AYA2016-77237-C3-3-P) and the Generalitat Valenciana (grant PROMETEOII/2014/069). Computer simulations have been carried out in the Red Espa\~nola de Supercomputaci\'on (Mare Nostrum and Tirant supercomputers) and in the Servei d'Inform\`atica de la Universitat de Val\`encia. I want to thank the organizers of the BHCB2018 meeting for their invitation, and specially Dr. Rita de Cassia Dos Anjos for her kind help during the celebration of the conference.}\\
        Departament d'Astronomia i Astrof\'{\i}sica, Universitat de Val\`encia, C/ Dr. Moliner, 50, 46100, Burjassot, Valencian Country, Spain.\\
        E-mail: \email{manel.perucho@valencia.edu}}

\abstract{There is compelling evidence showing that extragalactic jets are a crucial ingredient in the evolution of host galaxies and their environments. Extragalactic jets are well collimated and relativistic, both in terms of thermodynamics and kinematics at sub-parsec and parsec scales. They generate strong shocks in the ambient medium, associated with observed hotspots in FRII radio galaxies, and carve cavities that are filled with the shocked jet flow, dragging a large fraction of the interstellar gas along, in the form of slow, massive outflows within the host galaxies. In this paper, I discuss relevant processes associated to jet evolution in the frame of FRI-FRII dichotomy. In particular, I focus on the role of 1) the interaction between galactic atmospheres and the jet head on global FRII jet kinematics, and 2) mass load by stellar winds or small-scale instabilities on jet deceleration in FRI jets. The results presented are based on 3D relativistic hydrodynamical (RHD) and/or 2D axisymmetric, time-independent relativistic magnetohydrodynamical (RMHD) simulations.}

\FullConference{International Conference on Black Holes as Cosmic Batteries: UHECRs and Multimessenger Astronomy - BHCB2018\\
		12-15 September, 2018\\
		Foz du Iguazu, Brasil}

\begin{document}

\section{Introduction}

   Relativistic jets from active galactic nuclei (AGN) form in the surroundings of supermassive black holes (SMBH) \cite{bz77}. They present a well-know observational dichotomy \cite{fr74} that divides them into Fanaroff-Riley type II sources (FRII), which are bright, mainly at their edges, where they show the presence of strong hot-spots at the end of well collimated jets, and Fanaroff-Riley type I sources (FRI), less luminous in radio, with decollimated, fairly symmetric, jets at kiloparsec scales.
   
   Hotspots in FRIIs are understood as the impact site of the jet plasma in the ambient medium, where a reverse-shock in the jet reference frame is formed. The jet is thus necessarily supersonic at this point, most probably preserving mildly relativistic speeds at hundreds of kiloparsecs, as suggested by the jet to counter-jet brightness asymmetry produced by Doppler boosting (e.g., \cite{cb96,mh09}). 
   
   In the case of FRI jets, the brightness symmetry is explained by means of jet deceleration reducing the Doppler boosting. Deceleration takes place within the inner kiloparsecs since the formation region, i.e., well inside the host galaxy (e.g., \cite{lb14} and references therein). Several mechanisms have been proposed to explain FRI jet deceleration, involving mass load of interstellar medium (ISM) gas \cite{bi84}, interaction with clouds and/or stellar winds \cite{ko94,bo96}, via turbulent mixing. A way to entrain ISM gas is by means of the growth of instabilities. Entrainment and deceleration are probably related to a strongly dissipative process taking place on small scales that causes an increase of the jet's X-ray brightness \cite{lb14,kh12}. The lack of large-scale pinching or helical patterns probably excludes long wavelength, disrupting modes \cite{pm07,ro08}. Small-scale instability modes have been suggested in references \cite{pe10,ma17,gk18a,gk18b} in the context of large scale jet expansion and recollimation. Nevertheless, no hints of strong recollimation shocks are observed in a number of FRI's \cite{lb14}, as expected from large scale oscillations of the jet surface \cite{pm07} that would trigger the small-scale instability modes.
   From an analytical perspective, a relation between jet power and mean ambient medium density has been envisaged to directly influence jet evolution and the FRI-FRII dichotomy ($L_j/\bar{n}_a \sim 10^{44 - 45} \, {\rm erg s^{-1} cm^3}$ \cite{ka09}), which has been later confirmed by numerical simulations \cite{tch16}. The efficiency of mass-load by stellar winds depends on the stellar population and jet power.
      
   In this contribution, I review results based on both recently published and ongoing work on numerical simulations of relativistic flows in the context of radio source evolution and the complex dance they play with the host galaxy. In Section~\ref{s:frii}, I focus on the role of small amplitude of the jet head on FRII jet evolution, plus the triggering of massive outflows associated to these powerful jets. Section~\ref{s:fri} includes a summary of the results derived from ongoing work on the role of mass-load by stellar winds in FRI jets. Finally, Section~\ref{s:disc} collects some final remarks and conclusions. 

\section{FRII jets} \label{s:frii}

    As stated above, FRII jets remain collimated for very large distances. Among the factors related to the small jet opening angles we find the possible role of a toroidal field, cocoon pressure, and self collimation $\propto 1/\Gamma$ (where $\Gamma$ is the Lorentz factor). Although there is little information about the relevance of a toroidal field at kpc scales and polarization measures seem to indicate that the field is predominantly poloidal in FRIIs, we know from simulations that 1) the toroidal field can indeed help collimating FRII jets \cite{mi10}, 2) cocoon pressure is proportional to the injected energy \cite{pmqb17}, which is larger for the more powerful jets, and 3) the Lorentz factor is $> 1$ at hundreds of kiloparsecs from the SMBH \cite{mh09}.

\begin{figure*} 
\begin{center}
	\includegraphics[clip, trim=1.cm 5cm 1.cm 5cm,width=0.8\textwidth]{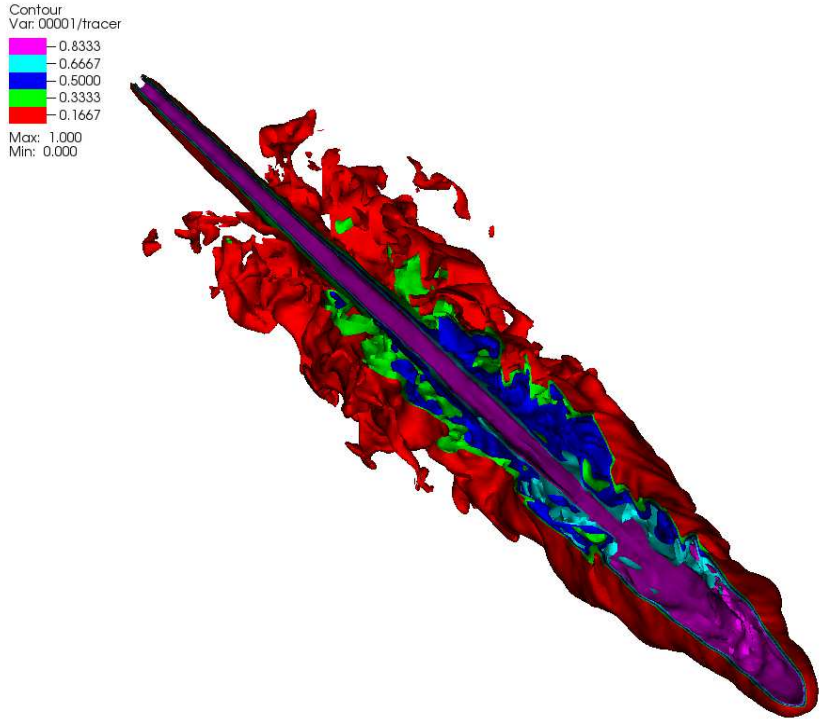}
    \caption{Jet mass fraction iso-surfaces of a relativistic jet propagating through a decreasing density ambient medium. The box is 200~kpc long in the jet direction.}
    \label{fig:trac}
    \end{center}
\end{figure*}

  Reference \cite{pmq18} includes 3D simulations of jet evolution through an ambient medium with a galaxy/group density profile, using a numerical box of $100\,{\rm kpc} \times 100\,{\rm kpc} \times 200\,{\rm kpc}$. These simulations follow previous axisymmetric simulations \cite{pqm11,pmqr14} aimed to study the role of relativistic outflows on the ISM and intergalactic medium (IGM). Both 2D and 3D simulations show that shock-driven heating of the ISM/IGM is very efficient, mainly due to an efficient conversion of the jet energy flux into cocoon (cavity) pressure at the reverse shock (hotspot). Actually, a large portion of the injected energy through the jet channel is transferred to the ambient medium. In reference \cite{pmqb17}, the authors showed that this is directly related to the relativistic nature of AGN jets. The concentration of high energy and momentum fluxes through a narrow channel favours high post-shock pressure:
 \begin{equation}\label{eq:p_Lrel}
 P_{h} \sim \frac{L_{j} \, v_{j}}{A_{j} \, c^2},
\end{equation}
with $P_h$ the hotspot pressure, $L_j$, $v_j$ and $A_j$ the jet power, velocity and cross-section at the reverse shock, respectively. Compared to typical values used in non-relativistic jet simulations, the authors showed that hotspot pressure can be more than one order of magnitude larger in the case of relativistic jets:
\begin{equation}
\frac{P_{h,c}}{P_{h,r}} \simeq 7 \times 10^{-2} \,
\left(\frac{\rho_{j,c}}{\rho_{j,r}}\right) \left(
\frac{c}{v_{j,r}}\right)^2 \left(\frac{0.3c}{v_{j,c}}\right)^{-2}
\left(\frac{c^2}{h_{j,r}\, \Gamma_{j,r}^2}\right),
\label{eq:pcpr}
\end{equation}
with $c$ and $r$ subscripts referring to classical and relativistic jets.
   
   Furthermore, 3D RHD simulations show that FRII jets can undergo an acceleration phase as the jet head propagates down the galaxy negative density/pressure gradient, not only because of the drop in ambient density, but also if a small amplitude helical perturbation is set at injection \cite{pmq18}. This effect was previously noted by \cite{al00} for short scale 3D RHD simulations of jets evolving in homogeneous media, and its relevance has been now confirmed by \cite{pmq18} for a decreasing density ambient medium. The cause of the enhanced acceleration is the obliquity of the reverse shock: In axisymmetric flows the reverse shock is typically a planar Mach disk, which implies strong head deceleration, whereas this effect is reduced in the case of oblique shocks. As a result, the jet head propagates much faster than expected from 2D axisymmetric simulations, as long as the amplitude of the perturbation remains linear. If the perturbation couples to a Kelvin-Helmholtz instability mode, its amplitude grows with distance and the jet eventually enters into the dentist-drill phase \cite{sch74}, in which the jet momentum is spread over a wide cross section and the jet head decelerates. 
   
    Some large-scale FRII jets show a remarkable straight trajectory (e.g., 3C 219, 3C 273), whereas others show kinks or bends at large scales (e.g., Cygnus A). In the picture described by \cite{pmq18}, the former would be observed within the initial, small amplitude oscillation phase, while the latter undergo large amplitude oscillations and are in the dentist-drill phase. Figure~\ref{fig:trac} shows a 3D contour image of jet mass-fraction to illustrate the large-scale jet structure at the end of the simulation. The jet mass fraction is an advected variable in our numerical code (see \cite{pe10} and references therein) that is 1 for pure jet material and 0 for the ambient gas, so intermediate values show mixing. Transversal cuts performed to the image allow to see the jet/cocoon structure.
        
     A comparison of the estimated jet ages via synchrotron cooling typically result in larger jet ages than the $\sim 5\times10^6\,{\rm yr}$ that the simulated jet needs to reach $\simeq 200\,{\rm kpc}$ (see Table 2 in \cite{pmq18} and references therein). Synchrotron aging times typically incorporate the assumption of equipartition between the gas and the magnetic pressures. This difference in the age of the radio source can be explained either because the magnetic field is over equipartition, or because of a delayed instability growth to the nonlinear regime caused by a very small perturbation amplitude at injection plus the numerical viscosity caused by the relatively low resolution in our simulations (see, e.g., \cite{pe04}). A faster growth would bring the jet head to nonlinear oscillations and deceleration earlier, thus delaying its advance and increasing the time to reach the end of the grid.

\begin{figure*} 
\begin{center}
	\includegraphics[clip, trim=1.cm 5cm 1.cm 5cm,width=0.45\textwidth]{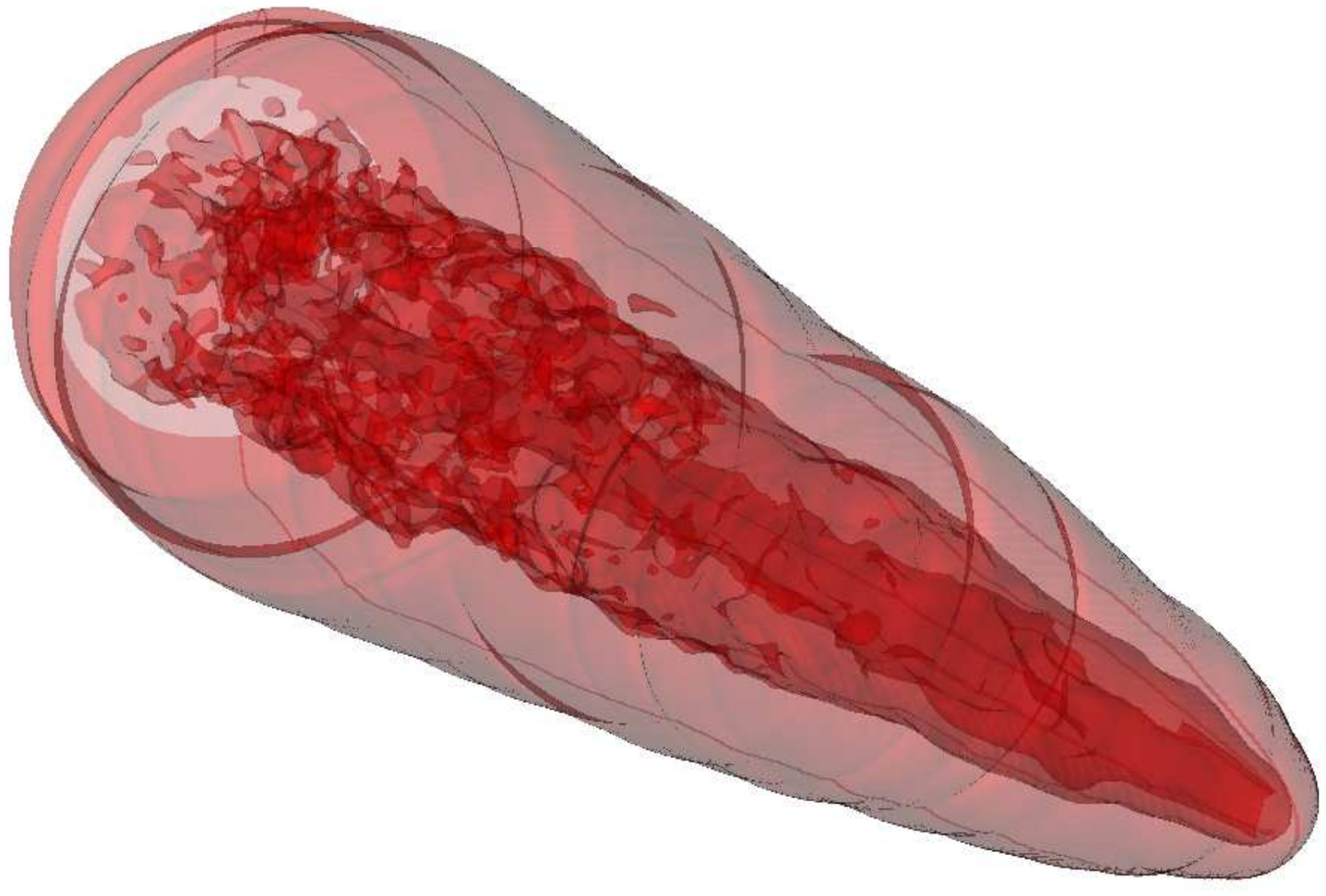} \includegraphics[clip, trim=1.cm 5cm 1.cm 5cm,width=0.45\textwidth]{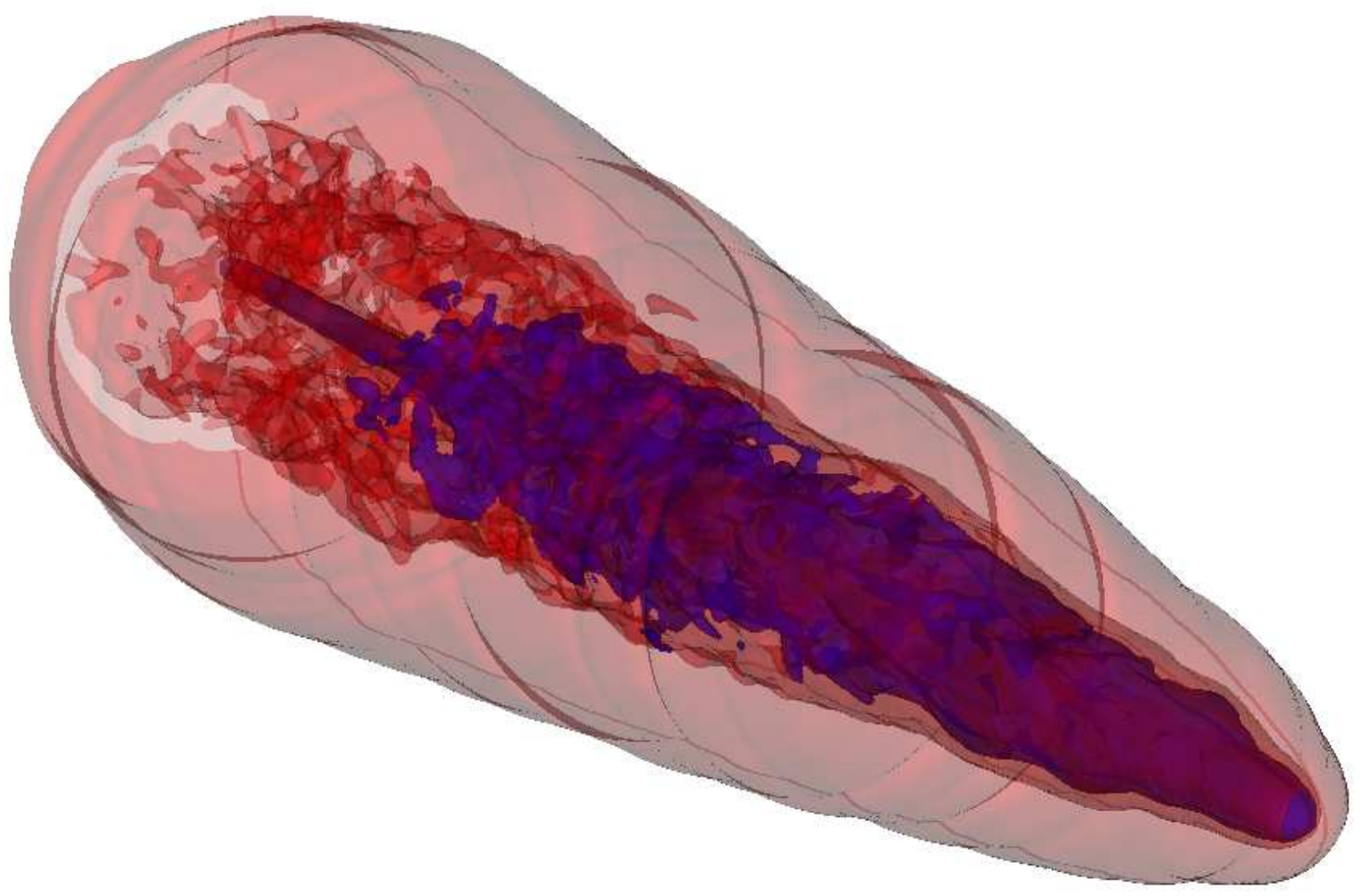}
    \caption{Left panel: Axial velocity isosurfaces at $100\,{\rm km/s}$. Transparency has been given to the surfaces to allow seeing the velocity structure of the shocked region through the bow-shock. Right panel: Overimpression of a tracer 0.5 isosurface.}
    \end{center}
    \label{fig:clou}
\end{figure*}

    Within the galaxy, the shock induced by jet expansion causes outward propagation of the ISM. The left panel of Figure~\ref{fig:clou} shows isosurfaces of axial velocity at $100\,{\rm km/s}$ (we have added transparency to allow the vision of the motions inside the bow-shock). The right panel includes an isosurface for jet-mass fraction $0.5$ (all cells outside this surface have less than 50\% of shocked jet gas) to show that these motions incorporate originally steady ambient medium. The image shows a region with a volume of $50\,{\rm kpc^3}$, so the shocked gas blobs propagating outwards are located at $\sim 20 - 40\,{\rm kpc}$ from the SMBH, i.e., within the host galaxy. In fact, massive outflows are observed at tens of kpc from the active nuclei associated to relativistic outflows (see, e.g., \cite{mo13} and references therein).

\section{FRI jets} \label{s:fri}

   As explained in the Introduction, the modellisation of FRI jet deceleration from observations implies a continuous and dissipative process \cite{lb14}. The main candidates to produce such effect are small scale instability modes that trigger turbulent mixing, or mass load by stellar winds and clouds. In this paper, I summarise recent results on mass load by stars. 
   
   The interaction between jets and stars can represent an efficient way to locally dissipate the jet kinetic energy as has been shown in, e.g., \cite{br12,pe17}. Taking into account that it has been shown that a large number of stars may be present inside the jet at a time \cite{ara13,wyk13,wyk15}, the global effect of these interactions may play a crucial role in jet evolution. Komissarov \cite{ko94} showed that the jet-star interaction could be treated as a hydrodynamical problem, because the Larmor radius of the particles is orders of magnitude smaller than the typical size of the interaction region. This scenario was later proposed as a plausible one to explain high energy emission from blazars and radiogalaxies \cite{bp97,bba10}. In \cite{br12,pe17} the authors have focused on the nature and dynamical effect of the jet-wind collision at different locations along the jet (few parsecs and $100\, {\rm pc}$, respectively). Their results indicate that the shock heating of the wind gas bubble/cloud can increase the cross-section of the obstacle, producing a large region of kinetic energy dissipation within the jet. Furthermore, three-dimensional simulations show that the cometary tail that forms immediately after the collision is strongly unstable and thus triggers strong mixing in small scales. Finally, this picture suggests a high degree of inhomogeneity in the jet caused by the discrete locations of jet/obstacle interactions. This is possibly observed in the nearby AGN jet of Centaurus A (e.g., \cite{wo08}).
   
  Bowman et al. \cite{bo96} studied the global effect of stellar mass-load in the case of low power, relativistic hydro jets using steady state models. They found that mass load by a typical stellar population of an elliptical galaxy composed of old, low mass stars could be enough to decelerate those low-power jets. An interesting dichotomy was pointed out: colder and faster jets can heat up in the process due to dissipation, whereas initially hotter jets cool down during the entrainment of the cold wind particles. Perucho et al. \cite{pmlh14} have run dynamical simulations of jet evolution including a source term for mass-load by a spherically symmetric distribution of stars that follow a Nuker profile and confirmed Bowman et al.'s results. They also showed that a stellar population with a typical wind of $\dot{M} = 10^{-12}\,{\rm M_\odot yr^{-1}}$ would be insufficient to explain jet deceleration in the case of FRI jets with powers $\geq 10^{43}\,{\rm erg/s}$. From an analytical perspective, in reference \cite{hb06} the authors estimate the mass-load required to efficiently decelerate a jet as a function of its power and show that a high mass-loss star $\dot{M} = 10^{-5}\,{\rm M_\odot yr^{-1}}$ alone could be enough to eventually decelerate a low power jet ($\simeq 10^{42}\,{\rm erg/s}$) while the star crosses the jet.  
   
   Ongoing work by this author and collaborators (Perucho, Mart\'{\i}, Angl\'es and Laing, in preparation) focuses on steady state solutions for leptonic, magnetised jets with power $10^{43}\,{\rm erg/s}$. Based on previous work, we decided to study the role of stellar populations with mass-loss in the range $10^{-9}\, -\, 10^{-11}\,{\rm M_\odot yr^{-1}}$. These mass-losses correspond red giant populations that are relevant or even dominant at the center of the host galaxy.
   
   The simulations are run with a one-dimensional code that returns the equilibrium structure for a jet under several assumptions, including that the jet velocity, $v^z$, is close to the speed of light and that the jet axial coordinate, $z$, is always much larger than the radial one, $r$, among others (see \cite{ko15,fue18} for details). Under these assumptions, the axial coordinate can be taken as the temporal one, and radial jet equilibrium is solved at each jet axial position (equivalent to a given time $t=z/c$). The steady jets are simulated from 10 to 500~pc, and are injected with a purely electron/positron composition, whereas the stellar wind is modeled as a source term of a cold electron/proton plasma (negligible energy flux) at rest (negligible momentum flux) in the mass equation.
   
   We have run many models ranging several orders of magnitude in rest-mass density, internal energy density, and three different Lorentz factors ($6-10$). The initial magnetic field configuration \cite{li89,mar15} together with the transversal equilibrium imposed do not allow magnetically dominated jets, but the magnetic to gas pressure ratios, $\beta$, in our models ranges from $0.1$ to $10$.  The results are highly dependent on the initial energy distribution of the jet, namely, the proportion of the energy flux in the form of kinetic, internal or magnetic energies. In general terms, initially kinetically dominated, denser jets require strong mass-loads to be decelerated ($10^{-9}\,{\rm M_\odot\,yr^{-1}}$). Otherwise, for milder values of stellar mass-loss ($10^{-11}\,{\rm M_\odot\,yr^{-1}}$), jet composition is altered and kinetic energy is dissipated, causing a slight increase of the flow internal energy with distance. In the case of internal energy dominated, dilute jets, even milder values of mass-loss completely change the jet composition and force a strong drop in the jet internal energy, albeit with minor deceleration. Stronger mean mass-loss rates efficiently decelerate these jets in distances $< 100\,{\rm pc}$. The different jets have been simulated for three different magnetic field configurations, namely for pitch angles of $10^\circ,\,45^\circ$ and $75^\circ$, and also without magnetic field (RHD jets). Interestingly, magnetised jets show smaller opening angles because of magnetic tension. This has a double effect on jet evolution: on the one hand, small opening angles avoid the penetration of further stars (or clouds) into the jet, and, on the other hand, if the jet has a large internal energy reservoir, extend the Bernoulli effect (thermal acceleration) along larger distances. Both effects delay jet deceleration while the toroidal field is significant. Furthermore, in cases in which the jet is strongly decelerated, the opening angle increases with distance, and we observe a drop in the magnetic energy flux that also extends the deceleration distance, i.e., also delays deceleration with respect to non-magnetised jets. 
      
\begin{figure*} 
	\includegraphics[width=\textwidth]{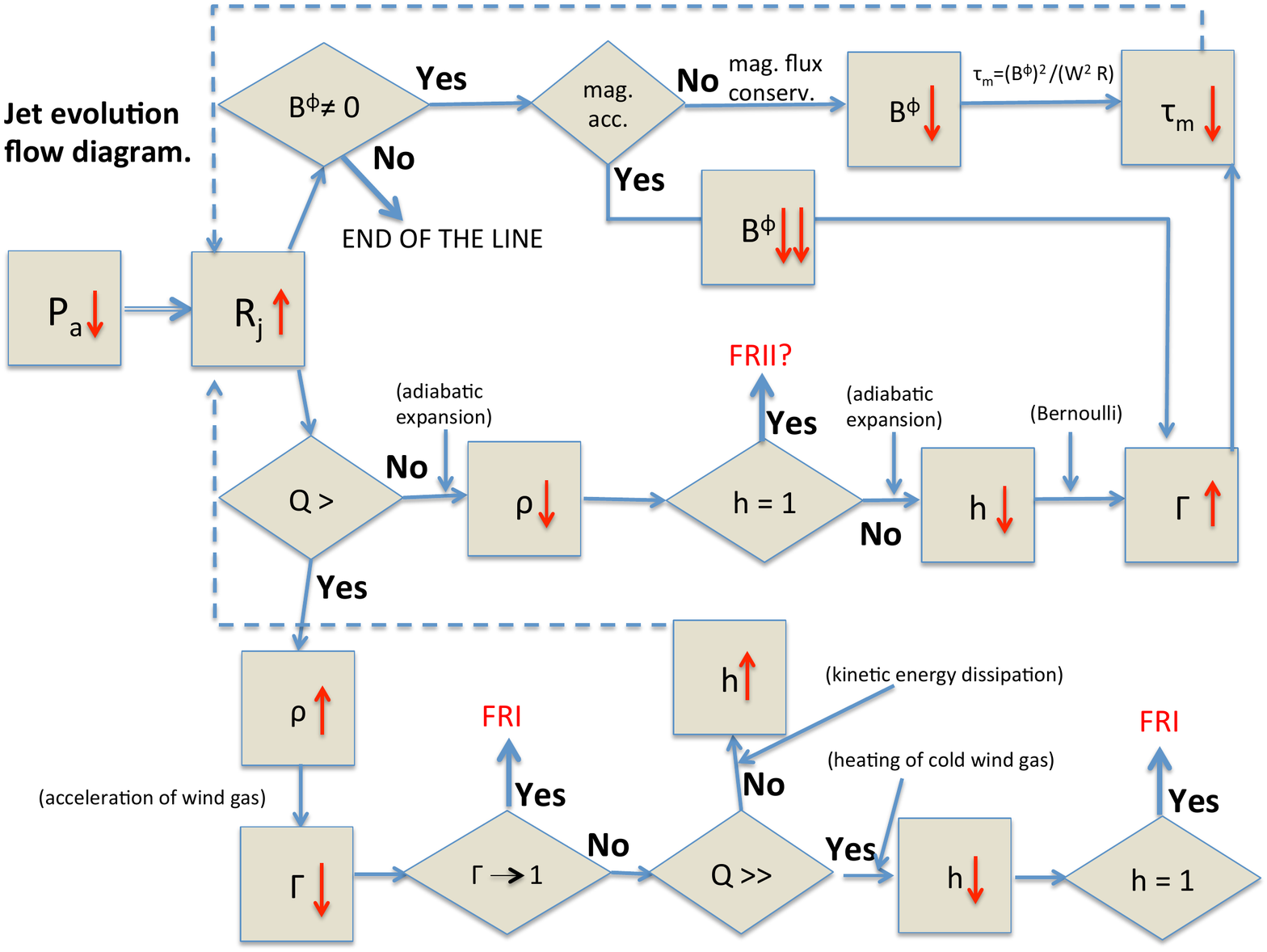}
    \caption{Flow diagram of jet evolution. The rhombi indicate bifurcations of the flow that depend on the local jet properties, whereas the squares indicate natural consequences of the described processes.}
    \label{fig:flow}
\end{figure*}

   The results of this work are summarised in Fig.~\ref{fig:flow}, which shows a flow diagram covering the different possibilities found in our wide parameter range. The diagram splits the jet evolution in a small number of discrete questions affecting key parameters, separately. Nevertheless, several processes may occur at the same time. For instance, although the first question is whether the jet is or not magnetised, on the one hand, and whether the mass-load it undergoes is relevant or not, on the other, these questions may need to be answered at the same time because they occur independently. 
    
     The flow diagram is based on basic conservation laws. Jet expansion with distance is common to all extragalactic jets, and it is caused by jet overpressure with respect to the ambient medium, e.g., when the ambient pressure drops. The opening angle is determined by the jet axial velocity and sound speed (Mach angle). In the case of magnetised jets with significant toroidal components, the expansion can be refrained by the magnetic tension ($\tau_m$ in the diagram). However, the conservation of magnetic flux forces a drop in the magnetic field inversely proportional to the jet radius ($\propto 1/R$) that weakens magnetic tension with distance. It is relevant to point out that this also depends on the opening angle of the jet: the larger the opening angle, the faster the drop of magnetic tension. In the case of differential expansion, probably associated to non-conical jet shapes, magnetic energy flux can be converted into kinetic energy (e.g., \cite{ko12} and references therein) and the magnetic field drops faster than $1/R$, which can contribute to weakening magnetic tension. 
                  
  Mass conservation relates the mass flux in the jet at $z_1$ with the injected one at $z_0$: 
\begin{equation} 
  \rho \Gamma A v\, = \, \rho_0 \Gamma_0 A_0^2 c\, + \int_{z_0}^{z_1} Q(z) A(z) dz,
\end{equation}
where subscript $\rho$ is the flow rest-mass density, $\Gamma$ is the Lorentz factor, $v$ is the velocity, $A$ is the jet cross-section, subscript $0$ refers to values at injection (where we have taken $v\sim c$), and $z$ is the axial coordinate. This relation allows us to estimate a minimum value for the mass-load $Q$ that makes the second term on the r.h.s. relevant. If $\int_{z_0}^{z_1} Q(z) A(z) dz << \rho_0 \Gamma_0 A_0^2 c$ (negligible mass-load), jet expansion causes a drop in rest-mass density and thermal acceleration (if the enthalpy $h$ is larger than one, i.e., if the jet flow carries a relevant budget of internal energy density). Otherwise, or as the internal energy reservoir decreases, the jet propagates at its terminal velocity, probably as an FRII. At kpc scales, dissipation caused by waves and instabilities can reduce the FRII jet Lorentz factor to smaller values than those estimated at pc scales.

  On the contrary, if $\int_{z_0}^{z_1} Q(z) A(z) dz \geq \rho_0 \Gamma_0 A_0^2 c$ (where mass-load can be provided by any source, e.g, stellar-winds or instability growth), we can expect an increase of the jet rest-mass-density and a resulting drop in jet axial velocity. If this process brings the flow to $\Gamma \simeq 1$ and transonic speeds, an FRI jet is produced. From the momentum conservation equation
  \begin{equation}
\pi R_0^2 \left( \rho_0 h_0\,\Gamma_0^2 \,c^2 \,+ P_0\right) \,=\, \pi R^2 \left( \rho h\,\Gamma^2 \,c^2 \,+ P\right).
\end{equation}    
and neglecting pressure on both sides of the equation, we can deduce that, as long as the jet keeps a relativistic velocity ($\Gamma > 1$), another option arises: if the kinetic energy reservoir in the jet is large enough and entrainment is such that dissipation can heat the resulting flow, there is an increase of internal energy/enthalpy; on the contrary, if the entrainment is so strong that the dissipation of kinetic energy is not sufficient to heat it, the internal energy of the jet decreases (see Perucho et al., in preparation, for details). On the one hand, deceleration plus cooling also lead to an FRI like jet, on the other hand, being the mass-load dynamically relevant, heating enhances jet expansion and acceleration, but also further entrainment and thus deceleration, so the same fate is probable. 

 It is worth mentioning that in cases in which mass load is relatively important, expansion has a runaway effect caused by the entrainment of more stars (or clouds) as the jet cross section grows. Finally, it is relevant to stress that in this flow diagram we describe competing effects that can take place at the same time for a given jet and can either have opposite consequences (and therefore compete), or contribute to the same effect.  
       

\section{Final remarks} \label{s:disc}

  Although the Fanaroff-Riley dichotomy of extragalactic radio sources is known to be directly related to the jet injection power \cite{rs91}, the evolution of both types of jets show the detailed interplay between jets and their host galaxies. In the case of powerful FRII jets, the jet suffers little dynamical influence from the ambient medium, but a few million years of jet injection condition the galaxy evolution for hundreds of millions of years, at the least. As we approach the bordering power of $\sim 10^{44}\,{\rm erg/s}$ between the two classes, the ambient medium starts playing a role. This is the energy range in which we could expect that, e.g., an asymmetry in gas density or eventual penetration of massive stars, produced a hybrid morphology radio source \cite{gkw00}, with jet and counter-jet showing opposed behaviours. At lower powers, the mean stellar population and number of embedded stars can be critical for jet deceleration. 
  
  From the results summarised in this paper, currently favoured scenarios to explain jet deceleration are either the growth of small scale instabilities \cite{ma17,gk18a,gk18b}, stellar mass-load (e.g., \cite{bo96,pmlh14}), or a combination of both. In reference \cite{lb14}, the authors show that deceleration seems to occur from the boundaries to the axis in some cases, and more homogeneously in other sources. The former deceleration pattern favours a growing mixing layer produced by small scale instabilities, whereas the latter could well be explained by stellar mass-load. From our work, we can state that mass-load by stellar winds is probably negligible in galaxies dominated by a low-mass, old population in combination with jet powers $\geq 10^{43}\,{\rm erg\,s^{-1}}$. On the contrary, it can play a fundamental role in jet deceleration if the mean stellar mass-loss rates are $\geq 10^{-10}\,{\rm M_\odot\,yr^{-1}}$, i.e., if the number of red giants is large enough at the inner kpc of the host galaxy, for the same jet powers. A detailed study of the stellar population in nearby AGN galaxies can thus clarify the role of stellar wind mass-load on jets (see, e.g., \cite{wyk15}).
  
  The process of mass-load, even if dynamically negligible, has a strong impact on jet composition and energetic distribution: The protons injected by the stellar winds may force jet 1) deceleration and cooling, in the case of low-power, fast jets, if the jet is internal energy dominated, triggering a transfer of internal energy into kinetic energy (due to increased density), 2) kinetic energy dissipation and heating, if the jet is kinetically dominated at injection, 3) magnetic acceleration, if the load induces differential expansion due to the increased gas pressure caused by a strong increase in particle number density. Moreover, only in cases in which the initial leptonic mass flux of the (kinetically dominated) jets is larger than the entrained mass flux, the jet remains predominantly pair dominated with distance. As a result, jet composition may not be a universal property far from injection (Perucho et al., in preparation).

\end{document}